\documentclass[sigconf]{acmart}
\usepackage{listings}
\usepackage{xcolor}
\AtBeginDocument{%
  }

\copyrightyear{2025} 
\acmYear{2025} 
\setcopyright{acmlicensed}\acmConference[GeoAI '25]{The 8th ACM
 SIGSPATIAL International Workshop on AI for Geographic Knowledge
 Discovery}{November 3--6, 2025}{Minneapolis, MN, USA}
\acmBooktitle{The 8th ACM SIGSPATIAL International Workshop on AI for
Geographic Knowledge Discovery (GeoAI '25), November 3--6, 2025,
Minneapolis, MN, USA}
\acmDOI{10.1145/3764912.3770813}
\acmISBN{979-8-4007-2179-3/2025/11}

\acmSubmissionID{123-A56-BU3}

\begin{document}

\title[AskNearby: LLM-Based Neighborhood Info Retrieval \& Recommendations]{AskNearby: An LLM-Based Application for Neighborhood Information Retrieval and Personalized Cognitive-Map Recommendations}

\author{Luyao Niu}
\authornote{Both authors contributed equally to this research.}

\author{Zhicheng Deng}
\authornotemark[1]
\affiliation{%
  \institution{Peking University}
  \institution{Qianmo Smart Link}
  \city{Shenzhen}
  \state{Guangdong}
  \country{China}
}
\email{luyao0160@stu.pku.edu.cn}
\email{zhicheng.deng@stu.pku.edu.cn}

\author{Boyang Li}
\authornote{Corresponding author.}
\affiliation{%
  \institution{New York University}
  % \institution{Qianmo Smart Link Tech}
  \city{Brooklyn}
  \state{NY}
  \country{USA}}
\email{boyang.li@nyu.edu}

\author{Nuoxian Huang}
\affiliation{%
  \institution{Imperial College London}
  % \institution{Qianmo Smart Link Tech}
  \city{London}
  \country{United Kingdom}}
\email{n.huang25@imperial.ac.uk}

\author{Ruiqi Liu}
\affiliation{%
  \institution{Tencent}
  \city{Shenzhen}
  \state{Guangdong}
  \country{China}}
\email{ruiqiliusysu@gmail.com}

\author{Wenjia Zhang}
\affiliation{%
  \institution{Tongji University}
  \city{Shanghai}
  \country{China}}
\email{wenjiazhang@tongji.edu.cn}

\renewcommand{\shortauthors}{Niu et al.}

\begin{abstract}
The "15-minute city" envisions neighborhoods where residents can meet daily needs via a short walk or bike ride. Realizing this vision requires not only physical proximity but also efficient and reliable access to information about nearby places, services, and events. Existing location-based systems, however, focus mainly on city-level tasks and neglect the spatial, temporal, and cognitive factors that shape localized decision-making. We conceptualize this gap as the Local Life Information Accessibility (LLIA) problem and introduce \textit{AskNearby}, an AI-driven community application that unifies retrieval and recommendation within the 15-minute life circle. \textit{AskNearby} integrates (i) a three-layer Retrieval-Augmented Generation (RAG) pipeline that synergizes graph-based, semantic-vector, and geographic retrieval with (ii) a cognitive-map model that encodes each user's neighborhood familiarity and preferences. Experiments on real-world community datasets demonstrate that \textit{AskNearby} significantly outperforms LLM-based and map-based baselines in retrieval accuracy and recommendation quality, achieving robust performance in spatiotemporal grounding and cognitive-aware ranking. Real-world deployments further validate its effectiveness. By addressing the LLIA challenge, \textit{AskNearby} empowers residents to more effectively discover local resources, plan daily activities, and engage in community life.

\end{abstract}





%% These dates are used for tracking the submission. Not needed for the initial submission. Commented by Deng
% \received{27 August 2025}
% \received[revised]{27 August 2025}
% \received[accepted]{30 September 2025}

\maketitle

\section{Introduction}

%% background: 15-minute city
% introduce the research question
% The 15-minute city concept, which envisions residents fulfilling daily needs within a short walk or cycle within 15 minutes, is primarily studied through the lens of physical accessibility \cite{moreno2021introducing,abbiasov202415}. However, its digital counterpart-ensuring residents can access timely, relevant information about immediate surroundings—remains a critical, underexplored challenge \cite{khavarian202315}. Yet digital infrastructure alone does not guarantee such accessibility, raising a key question: how can people efficiently obtain local life information in dynamic environments where services, events, and places are constantly in flux?

The 15-minute city concept envisions that people can fulfill basic daily needs—such as grocery shopping, healthcare, education, and leisure—within a short walking or cycling distance from their homes, typically within 15 minutes \cite{moreno2021introducing}. This idea has gained widespread attention among urban planners and policymakers, who increasingly emphasize proximity-based planning and active transportation as essential principles for promoting sustainable urban development. In parallel with this, researchers have conducted empirical studies on whether residents can reach nearby destinations and access urban functions \cite{abbiasov202415,li2024enhancing,xu2025using, bruno2024universal}. 

Beyond physical accessibility, digitalization represents another core principle of the 15-minute city. It entails leveraging smart city technologies—such as big data and the Internet of Things—to support real-time information delivery, citizen engagement, and efficient urban resource management \cite{khavarian202315}. However, digital infrastructure alone does not guarantee that individuals can effectively access or utilize localized information \cite{donio2025neighborhood}. This raises a critical but largely unaddressed question: Can people access relevant information about nearby places and services to make informed decisions in their daily lives, especially in dynamic environments where places, facilities, and human activities are constantly changing?

%% Further characterization of local information
% Existing studies & research gap (traditional search/recommendation & LLMs)
Existing applications are ill-equipped to address this need of accessing local information at a granular, neighborhood scale \cite{meng2018,aliannejadi2019joint}. Their failure stems from an inability to model the nuanced spatiotemporal context, ensuring that the offered information is both timely and locally relevant. To overcome this limitation, it is necessary to support two complementary processes through which users interact with local information: \textit{active retrieval} and \textit{passive recommendation}.

In active retrieval, users explicitly seek information regarding nearby services or events, often formulating queries inherently containing spatiotemporal constraints (e.g., 'Which supermarket near me is open late?'). Traditional search engines rely heavily on keyword-based matching and fixed boundaries, such as a predefined search radius or time windows, failing to grasp the complex spatial and semantic intent in human queries \cite{zheng2011learning}. Recent studies have explored the potential of Large Language Models (LLMs) to interpret semantic and spatiotemporal constraints on local information and activities, but most approaches still treat space and time as isolated filters rather than as interdependent factors shaping local knowledge \cite{li2024urbangpt,yu2025spatial,ni2025tp}.

Passive recommendation proactively surfaces relevant information without explicit user queries. Achieving effective passive recommendation in a local context necessitates understanding not only individuals' preferences but also the broader spatial cognition that shapes their activity choices within a community. Although recommender systems have been extensively studied in urban computing \cite{zheng2014urban}, most existing approaches operate at a city-wide scale, relying on conventional collaborative filtering or content-based methods \cite{quijano2020recommender,aliannejadi2019joint}. These methods, however, often fail to accurately capture community-scale behavioral patterns or leverage residents’ unique cognitive understanding of places. As a result, they struggle to provide localized recommendations that align with how people perceive and navigate their immediate environments. 

%% Our contribution
% Problem Definition
To address these challenges, we define the problem of \textbf{Local Life Information Accessibility} (LLIA) as enabling residents to efficiently obtain relevant, timely, and context-aware local information through both active search and passive recommendation. Distinct from traditional search or recommendation tasks, LLIA requires jointly modeling spatial proximity, temporal relevance, and cognitive familiarity at the neighborhood scale, which poses unique technical and representational challenges. To support LLIA, we propose \textit{AskNearby}, an AI-driven community system built on a unified spatiotemporal retrieval and recommendation framework powered by LLMs. \textit{AskNearby} consists of two main components: First, a three-layer Retrieval-Augmented-Generation (RAG) architecture that facilitates active information seeking by synergizing \textbf{GraphRAG} (graph-based retrieval), \textbf{VectorRAG} (semantic similarity retrieval), and \textbf{GeoRAG} (geographic-based retrieval); second, a cognitive map model that captures users’ spatial preferences to support passive, location-aware personalized recommendations.

Our main contributions are threefold: 
\begin{itemize}
\item We define and formalize the LLIA problem, which encompasses both active search and passive recommendation to support spatiotemporal, context-aware access to neighborhood-scale information.
\item We design and develop \textit{AskNearby}, a novel community platform that integrates a multi-layered LLM-based RAG architecture to enhance local information retrieval and a cognitive map-based recommendation system.
\item We validate \textit{AskNearby} through extensive evaluations of \textit{AskNearby} on real-world local community datasets, demonstrating its superior performance in providing more accurate and context-aware local information compared to baseline methods.
\end{itemize}

\section{Related Works}
This section reviews prior research on spatiotemporal information retrieval and recommendation, with a particular focus on methods that leverage textual data, natural language queries, and user-generated content. We begin by reviewing traditional spatial and temporal approaches, and subsequently, we discuss recent advances driven by LLMs.

\subsection{Local Information Retrieval and Recommendation}
Recent work on local information retrieval and recommendation primarily addresses the spatial properties of Points of Interest (POIs) and the temporal signals captured in user-generated or activity-related content, such as check-ins, textual reviews, and event listings. A range of techniques have been developed to improve retrieval accuracy and recommendation efficiency. Representative approaches include context-aware modeling of user behavior \cite{aliannejadi2018personalized,zhao2022nextpoi}, spatial indexing and query optimization \cite{10.1145/3397271.3401090,meng2018}, time-sensitive recommendation strategies \cite{aliannejadi2019joint,shen2018time}, and hybrid systems that combine content-based and collaborative filtering \cite{Yochum2020LBR}. 

However, existing approaches tend to focus on retrieving structured, global-scale content rather than localized, contextualized knowledge. This leads to two key limitations. First, users often need to interpret search results to extract meaningful local insights manually. Second, most systems are not designed to handle open-domain, natural-language queries about local life—queries that are inherently ambiguous, context-rich, and highly personalized.

\subsection{LLMs for Spatiotemporal Information Retrieval and Recommendation}
To address these challenges, recent work has turned to LLMs, which have shown strong capabilities in understanding natural language queries and reasoning over complex content \cite{li2024urbangpt}. Empirical studies demonstrate that LLMs can effectively process time-series data, geospatial inputs, and crowd-sourced textual content \cite{han2025adapting,huang2025llm}. They are also capable of inferring temporal relations from unstructured text \cite{yu2023temporal}, improving geolocation representations through language understanding \cite{he2025geolocation}, and identifying implicit spatial references embedded within user expressions \cite{yin2025llm}. These emergent capabilities position LLMs as highly promising tools for spatiotemporal information access, particularly in local contexts where spatial, temporal, and cognitive factors are deeply intertwined \cite{li2024urbangpt}.

Capitalizing on these strengths, recent efforts have applied LLMs to spatiotemporal information retrieval and recommendation tasks. For example, EvoRAG integrates retrieved user trajectories with LLM reasoning to support personalized travel planning \cite{niTPRAGBenchmarkingRetrievalaugmented2025}. Tian et al. \cite{tianAdvancingLargeLanguage} present a retrieval and re-ranking framework that uses LLMs to identify similar environmental events from web and news sources. Li et al. \cite{liLargeLanguageModels2024} propose an LLM-based framework for next POI recommendation that preserves rich contextual signals from location-based social network data, effectively addressing challenges such as cold-start and short user trajectories. LLMs have also exhibited potential in interpreting nuanced spatial descriptions (e.g., "nearby", "within walking distance") \cite{haris2024exploring} and temporal phrases (e.g., "open tonight", "happening this weekend") \cite{yuan2024back}, which traditional systems often struggle to handle.

Despite these notable advances, current LLM-based applications typically lack a holistic spatiotemporal perspective \cite{LLMsPOI2024,RSLLMs2024}. Such systems often prioritize textual semantics while overlooking the complex interplay of time, space, user context, and cognitive spatial familiarity. Consequently, significant challenges remain in delivering user-centric, context-aware, and personalized access to local life information—especially within the dynamic, fine-grained environments envisioned by the 15-minute city paradigm.

\section{Methodology}
\subsection{Problem Definition: LLIA}
LLIA refers to the ability of residents to efficiently obtain timely and context-aware local information. It encompasses two complementary processes: active retrieval, which is based on explicit user queries, and passive recommendation, which is driven by users' spatial familiarity and local preference patterns.

Formally, a user $u$ is characterized with spatial position ($s_u$), temporal context ($t_u$), and frequently visited places ($p_u$), which approximate cognitive preferences. Given a spatiotemporal knowledge base $K$ (containing POIs, events, and user-generated content with geographic, temporal, and semantic metadata), LLIA aims to return a set of relevant items $I = \{i_1, \dots, i_n\}$ that are both recent and proximate. The LLIA function ($A$) can be defined as:
\begin{equation}
    A(u) = \underbrace{\text{Retrieve}(Q, s_u, K)}_{\text{Active retrieval}} \cup \underbrace{\text{Recommend}(s_u, t_u, p_u, K)}_{\text{Passive recommendation}}.
\end{equation}

The $\text{Retrieve}(Q, s_u, K)$ module retrieves candidate items relevant to the user's query $Q$ and current location $s_u$. The $\text{Recommend}(s_u, t_u, \\p_u, K)$ module further ranks or selects results based on cognitive relevance, incorporating semantic similarity, spatial proximity, and public familiarity derived from $s_u$, $t_u$, and $p_u$.

% For active retrieval, the retrieval process is defined as:
% \begin{equation}
%     \text{Retrieve}(Q, s_u, t_u, K) = \{ i \in K \mid \text{rank}_{\Phi(Q,s_u, t_u, \cdot)}(i) \leq k \},
% \end{equation}
% where $\Phi(Q, s_u, t_u, i)$ represents a relevance scoring function that evaluates how well an item $i$ in the knowledge base $K$ satisfies the constraints embedded in $Q$, and $k$ is the preset number of required items. This function may incorporate semantic similarity, spatial proximity, and temporal availability.

% For passive recommendation, the system selects a set of top-$k$ items $I'$ from the candidate pool $K$ based on the user's spatial cognitive preferences:
% \begin{equation}
%     \text{Recommend}(s_u, t_u, p_u, K) = \{ i \in K \mid \text{rank}_{\Psi(s_u, t_u, p_u, \cdot)}(i) \leq k \},
% \end{equation}
% where $\Psi(s_u, t_u, p_u, i)$ is a relevance scoring function that captures the user’s spatial cognitive preferences, incorporating semantic similarity, geographic proximity, and public familiarity.

\subsection{System Overview}
\begin{figure*}[h]
  \centering
  \includegraphics[width=\textwidth]{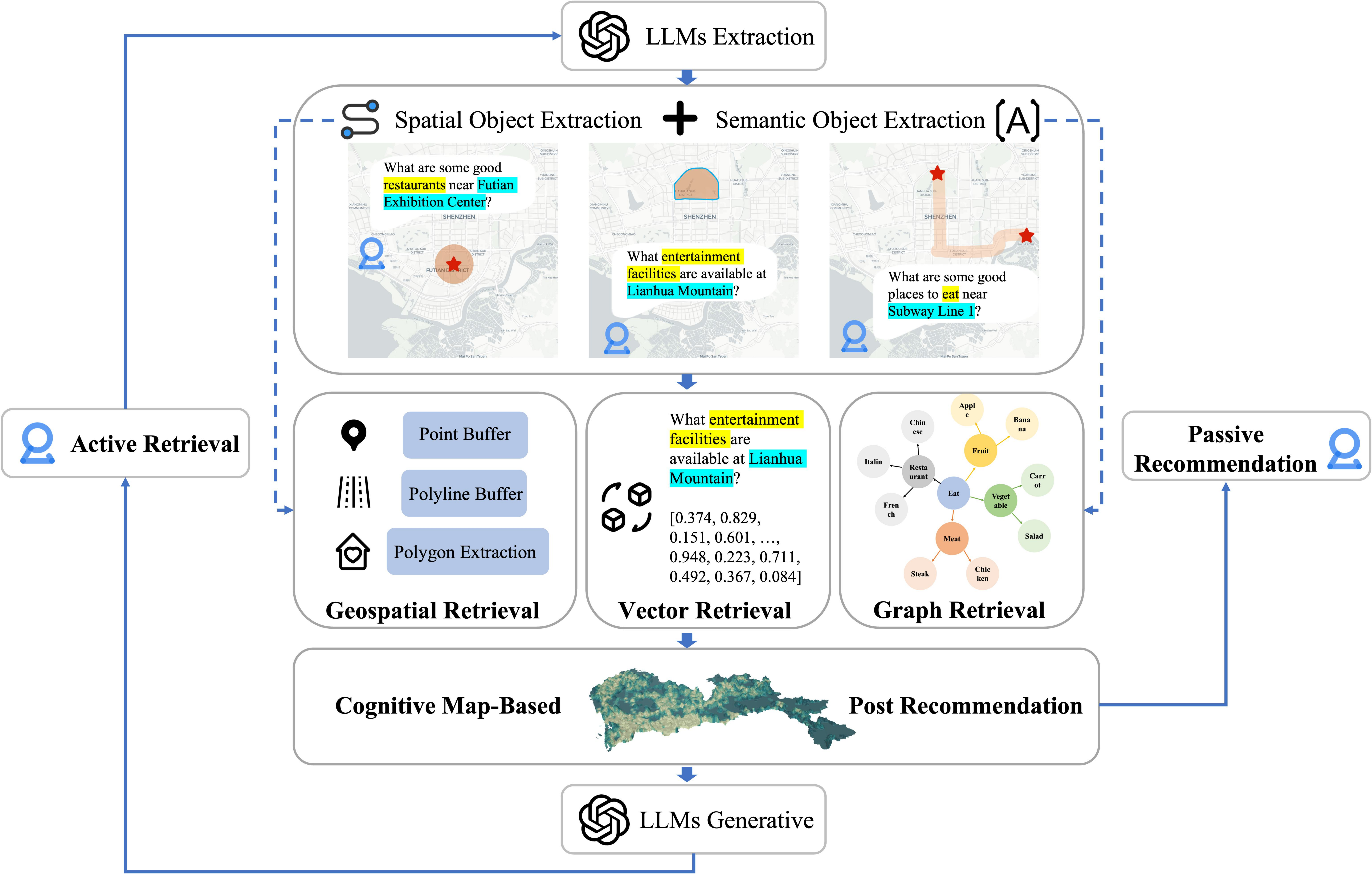}
  \caption{System overview of \textit{AskNearby}, combining LLM-based object extraction, multi-source retrieval (geospatial, vector, graph), and a cognitive map model to support active information retrieval and passive local recommendations.}
  \label{system_overview}
\end{figure*}

To enhance LLIA and provide local knowledge in a user-centric manner, we propose \textit{AskNearby}, an AI-driven community system. An overview of this system is illustrated in Figure \ref{system_overview}. \textit{AskNearby} combines a multi-layered RAG architecture with a cognitive map model that captures users’ spatial preferences. This unified framework enables both active information retrieval and passive, personalized recommendations based on spatial familiarity. Lastly, it provides an interactive user interface designed to facilitate real-time question-answering, personalized recommendation posts, and map-based visualization of nearby local information, as demonstrated in Figure~\ref{demonstration} of Appendix~\ref{apx:demonstration}.

\subsection{RAGs for Local Information Access}
To support open-domain, spatiotemporal local queries, we employ a multi-layer RAG framework that decomposes and enriches the user query via a structured processing pipeline. Upon receiving a user-issued query $Q$, an LLM is initially utilized to extract spatial entities (e.g., "Futian Exhibition Center") and semantic intents (e.g., "restaurants", "entertainment") from the input. 

Subsequently, a geospatial agent is invoked to query external APIs (e.g., Gaode Maps) and construct geometric representations of the identified locations in the form of points, polylines, or polygons. These structured spatial objects, together with the user’s current location, are passed to GeoRAG to retrieve geographically proximate candidates. 

In parallel, the semantic intent extracted from the query is used to direct GraphRAG, which retrieves conceptually or relationally relevant entities. Finally, the original query $Q$ is concatenated with the graph-expanded semantic results and sent to VectorRAG to retrieve semantically similar posts or documents in the vector space.

The overall retrieval pipeline is expressed as:
\begin{align}
\text{Retrieve}(Q,s_u) &= \text{GeoRAG}(\text{Loc}_{Q},s_u) \nonumber \\
&\quad + \text{GraphRAG}(\text{Sem}_{Q}) \nonumber \\
&\quad + \text{VectorRAG}(Q \oplus \text{Sem}_{G}),
\end{align}
where $\text{Loc}_{Q}$ and $\text{Sem}_{Q}$ denote location and semantic intent extracted from $Q$, and $\text{Sem}_{G}$ is the graph-augmented semantic context.

\textbf{GeoRAG} focuses on accurate spatial computation and location-based filtering. Backed by a spatial database (e.g., \textit{PostGIS}\footnote{\url{https://postgis.net/}}), it supports geo-indexing, proximity querying, and spatial joins. GeoRAG ensures that search results are geographically constrained and contextually relevant, such as identifying facilities within a specific walking distance or filtering nearby services based on real-time user location. Given a query location $\text{Loc}_{Q}$ (if available) or the user's current location $s_u$, and a candidate set of spatial entities $S$ that filtered from the spatiotemporal knowledge base $K$, GeoRAG retrieves:
\begin{equation}
\text{GeoRAG}(\text{Loc}_{Q}, s_u, S) = \left\{ s \in S \,\middle|\, \text{dist}(\text{Loc}, s) < \theta \right\},
\end{equation}
where $\text{Loc} = \text{Loc}_{Q}$ if provided, otherwise $\text{Loc} = s_u$. Here, $\text{dist}(\cdot)$ denotes a spatial distance function (e.g., haversine), and $\theta$ is a predefined threshold (e.g., 1km).

\textbf{GraphRAG} leverages semantic graphs to retrieve conceptually and relationally relevant information. Built on a graph database (e.g., \textit{NebulaGraph}\footnote{\url{https://www.nebula-graph.com.cn/}}), it supports efficient traversal, neighborhood expansion, and relation-based filtering. The semantic graph encodes relationships between entities such as POI categories, tags, events, or user interests. Given a query intent $\text{Sem}_{Q}$ extracted from the user input, GraphRAG identifies semantically related nodes and paths based on graph connectivity and relation types. For a graph $G$---constructed based on the information from $K$---and query intent $\text{Sem}_{Q}$, the semantic expansion can be defined as:
\begin{equation}
\text{GraphRAG}(\text{Sem}_{Q}, G) = \{ e \in G \mid \text{rel}(\text{Sem}_{Q}, e) \in R \},
\end{equation}
where $R$ denotes a set of predefined semantic or relational edges in the graph, and $\text{rel}(\cdot)$ indicates the presence of a meaningful semantic connection.

\textbf{VectorRAG} enables semantic-level retrieval by supporting approximate nearest neighbor search over candidate contents. VectorRAG refines results from the outputs of GeoRAG and GraphRAG by ranking them based on semantic similarity to the enhanced query. Specifically, we concatenate the original query \(Q\) with graph-augmented semantic context \(\text{Sem}_{G}\) to form an enriched input \( Q^{\prime} = Q \oplus \text{Sem}_G \).

We utilize \textit{pgvector}\footnote{\url{https://github.com/pgvector/pgvector}} as the vector similarity search engine, allowing efficient storage and querying of dense embeddings within a PostgreSQL-compatible environment. Both user queries and local content are embedded into high-dimensional vector representations through LLM-based encoders. Specifically, we adopt the \textit{bge-large-zh-v1.5} model released by the Beijing Academy of Artificial Intelligence \cite{bge_embedding}, which has demonstrated strong performance in semantic understanding tasks for Chinese text. We use Euclidean distance as the similarity metric to retrieve the most semantically relevant results from the filtered candidate set $V'$ composed of these outputs from GeoRAG and GraphRAG. Formally, with encoder $\phi(\cdot)$ and candidate set $V'$ derived from previous modules:
\begin{equation}
\text{VectorRAG}(Q, V') = \left\{ v \in V' \,\middle|\, \text{sim}\left(\phi(Q \oplus \text{Sem}_G), \phi(v)\right) > \delta \right\},
\end{equation}
where the function $\text{sim}(\cdot)$ denotes a similarity measure derived as the Euclidean distance between embeddings, and the threshold $\delta$ controls the minimum similarity required for a candidate $v$ to be selected.

Beyond active retrieval, personalized and context-aware recommendations are also essential for enhancing accessibility to local life information. To this end, we introduce a cognitive map model that captures individual user preferences across space and time, as described in the following subsection.

\subsection{Cognitive Map Model for Personalized Recommendation}
To support contextualized and personalized recommendations in local life scenarios, we propose a cognitive map model that captures how users internally perceive and associate meaning with urban spaces over time. Users' preferences vary not only by spatial proximity but also by how they functionally and socially interpret different places. To translate these human-centered cognitive preferences into actionable recommendations, we define the recommendation score for an information item $i$ as:
\begin{equation}
\Psi(s_u, t_u, p_u, i) = f_{\text{sem}}(s_u, t_u, p_u, i)^{\alpha} \cdot f_{\text{dist}}(d(s_u, i))^{\beta} \cdot f_{\text{pop}}(i)^{\gamma},
\end{equation}
where $\alpha,\beta,\gamma$ control the relative contributions of semantic relevance, spatial proximity, and public familiarity.

In this formulation, the overall recommendation score is computed as the product of three components: the \textbf{semantic relevance} term $f_{\text{sem}}(s_u, t_u, p_u, i)$, which measures the similarity between the user’s cognitive profile and the temporal functional semantics of the item’s location; the \textbf{spatial proximity} term \(f_{\text{dist}}(d(s_u, i))\), which models geographic relevance via a distance-decay function based on the spatial distance between the user and the item; and the \textbf{public familiarity} term $f_{\text{pop}}(i)$, which reflects the aggregated popularity or visitation frequency of the location, capturing its collective cognitive salience.

To reflect the relative importance of different components in the recommendation model, we assign each factor an exponent $\alpha$, $\beta$, and $\gamma$, which serve as soft weights in the multiplicative scoring function. We initialize these three weights equally at 1.0 to avoid imposing any prior preference, allowing semantic relevance, spatial proximity, and public familiarity to contribute equally to the final ranking. We acknowledge that a more optimal weight combination could potentially be found through grid search on the validation set, and we leave this for future work.

\begin{figure}[htbp]
  \centering
  \includegraphics[width=\linewidth]{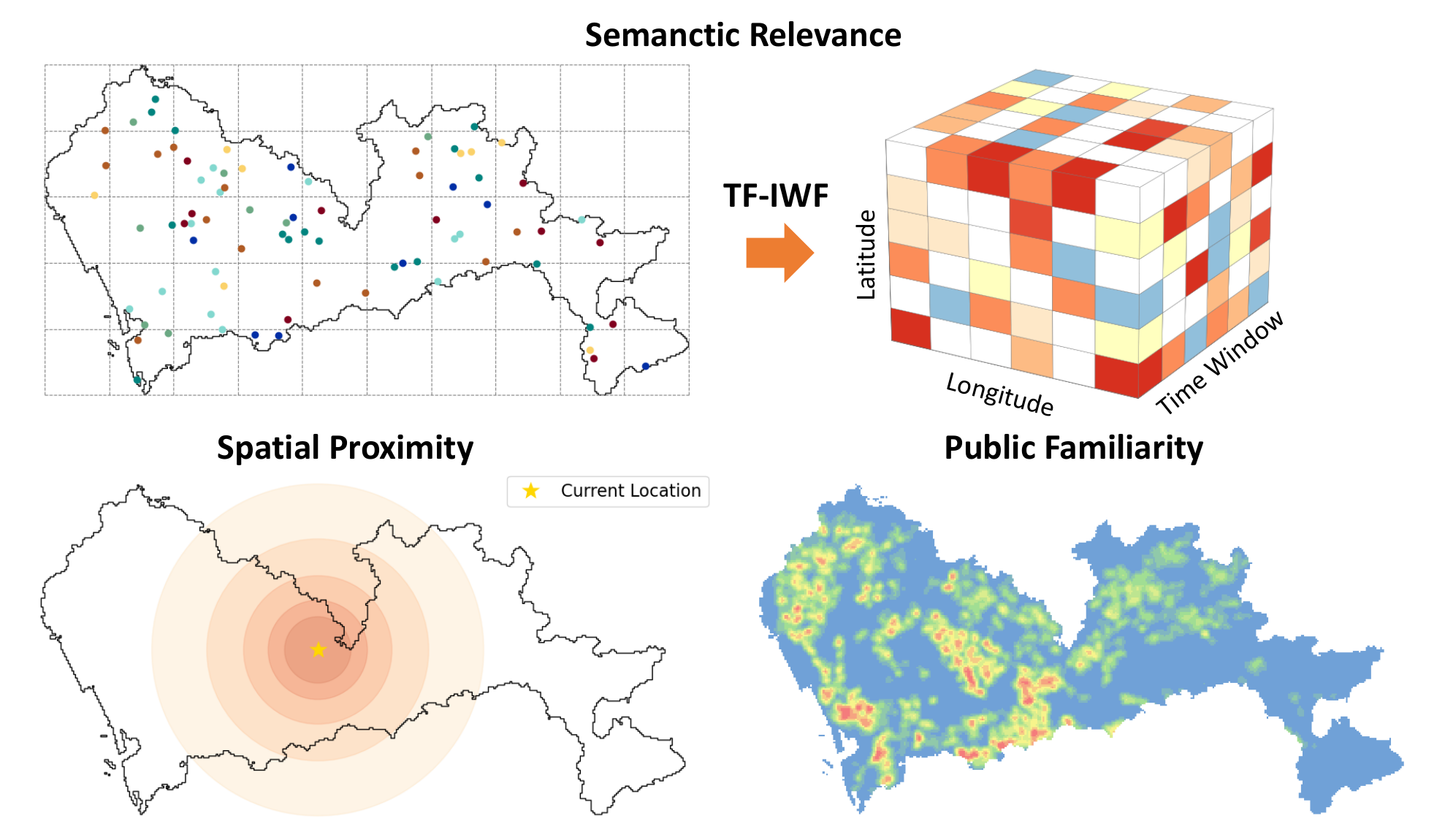}
  \caption{Illustration of the recommendation framework in \textit{AskNearby}, combining semantic relevance, spatial proximity, and public familiarity to compute personalized scores.}
  \label{fig:cognitive}
\end{figure}

Figure~\ref{fig:cognitive} illustrates the conceptual model behind this scoring framework, showing how the semantic, spatial, and popularity signals interact to determine which content to recommend to the user in different contexts.

% \textit{Semantic relevance.} We capture how users cognitively interpret places using TF-IWF-based vectors \cite{gui2021lsi}, which characterize the time-varying functional semantics of locations. The semantic score is computed by matching the item’s temporal semantics with the user’s cognitive profile, constructed from frequently visited places and the current context.

\textit{Semantic relevance.}
To model how users cognitively perceive and associate meaning with places, we implement the semantic component using a TF-IWF (Term Frequency-Inverse Word Frequency) based representation \cite{gui2021lsi}, which captures the time-varying functional semantics of each location. This scheme identifies place attributes that are locally prominent yet globally distinctive across different time periods. Formally, the TF-IWF score is computed as:
\begin{equation}
\text{TF-IWF}_{k,g} = \frac{A_{k,g}}{\sum_i A_{i,g}} \cdot \log\left(\frac{\sum_h A_{k,h}}{\sum_i \sum_h A_{i,h}}\right),
\label{tf-iwf}
\end{equation}

where $A_{k,g}$ denotes the count of functional attribute $k$ (e.g., restaurant, office, park) at location $g$ during a specific time window. For example, a location may exhibit office-related semantics during the day and shift toward dining in the evening. 

The semantic similarity $f_{\text{sem}}(s_u, t_u, p_u, i)$ is computed as the inner product between the TF-IWF vector of the item’s location at time $t_u$ and the user’s cognitive profile. This profile is constructed by aggregating the TF-IWF vectors of the locations in $p_u$, the set of frequently visited places by user $u$. Each location’s contribution is weighted based on its temporal proximity to the current time $t_u$. In addition, the TF-IWF vector of the user's currently location $s_u$ at time $t_u$ is included to capture short-term contextual relevance.

This design allows the semantic matching to reflect both the user’s long-term spatial familiarity and the immediate context of their current activity.

\textit{Spatial proximity.}
We model spatial relevance using a distance-decay function $f_{\text{dist}}(d) = \exp(-\lambda_d \cdot d(s_u, i))$, where $d(s_u, i)$ is the geographical distance between user and item, and $\lambda_d$ controls the decay rate. This formulation encourages recommending closer items while still allowing semantically relevant distant items to appear.

\textit{Public familiarity.}
To account for socially shared knowledge, $f_{\text{pop}}(i)$ reflects how frequently item $i$'s location has been visited by the general public. This component promotes widely recognized places and enhances collective cognitive relevance, thereby increasing the exposure of popular and trustworthy content in recommendations.

By integrating personalized semantics, spatial constraints, and collective familiarity, our model provides passive recommendations that are both user-centric and socially grounded. This allows the system to suggest nearby and timely content that reflects both individual interests and shared patterns in the local community.

% By integrating personalized semantics, spatial constraints, and collective familiarity, our model provides passive recommendations that are both user-centric and socially grounded. This allows the system to suggest nearby and timely content that reflects both individual interests and shared patterns in the local community.

\section{Experiments and Results}
\subsection{Data Description}
To facilitate a robust empirical evaluation of \textit{AskNearby} and its comparison with other LLM-based applications, we constructed a dataset of real-world, location-based user-generated content. This data was sourced from RedNote, one of China's foremost social media platforms for local discovery and lifestyle sharing. The study area is set in Shenzhen, a large metropolitan city in China. We programmatically collected geotagged posts associated with major landmarks (e.g., Nantou Ancient Town, Wutong Mountain) by targeting location-specific keywords. 

To mitigate potential biases from RedNote's native recommendation algorithm, we implemented a three-stage data collection and processing methodology. The process began with a broad keyword search across the study area to gather a comprehensive and unfiltered corpus of candidate posts. This initial corpus was then subjected to a rigorous cleaning phase, which included deduplication, the removal of advertisements and spam, and sample-based verification of geographic information accuracy. Finally, from this refined corpus, we performed random sampling to constitute our final dataset, thereby ensuring it is a representative, rather than a platform-biased, sample of user-generated content.

The resulting dataset comprises approximately 20,000 posts and encompasses a diverse range of popular activity areas, including historical sites like Nantou Ancient Town and natural attractions such as Wutong Mountain. This selection ensures comprehensive representation across a spectrum of urban functions, from commercial and recreational zones to key transportation hubs. Each record contains essential metadata such as post content, timestamp, user attributes (e.g., type), interaction metrics, semantic tags, and geographic coordinates. Table~\ref{tab:fields} provides a summary of the data schema.

\begin{table}[ht]
\small
\centering
\caption{Key fields in the RedNote dataset}
\label{tab:fields}
\begin{tabular}{ll}
\toprule
\textbf{Field} & \textbf{Description} \\
\midrule
Title, Content & Textual description of the post \\
Timestamp & Posting date and time \\
Author, Type & Username and whether verified or not \\
Likes, Comments & Interaction counts \\
Semantic Tags & Keywords, topics, sentiment labels \\
Location Name & Place name extracted or tagged \\
Longitude, Latitude & Geographic coordinates (WGS84) \\
Media Type & Post format (text, image, video) \\
Original or Repost & Indicates whether the post is original content \\
\bottomrule
\end{tabular}
\end{table}

\begin{figure}[!htbp]
  \centering
  \includegraphics[width=\linewidth]{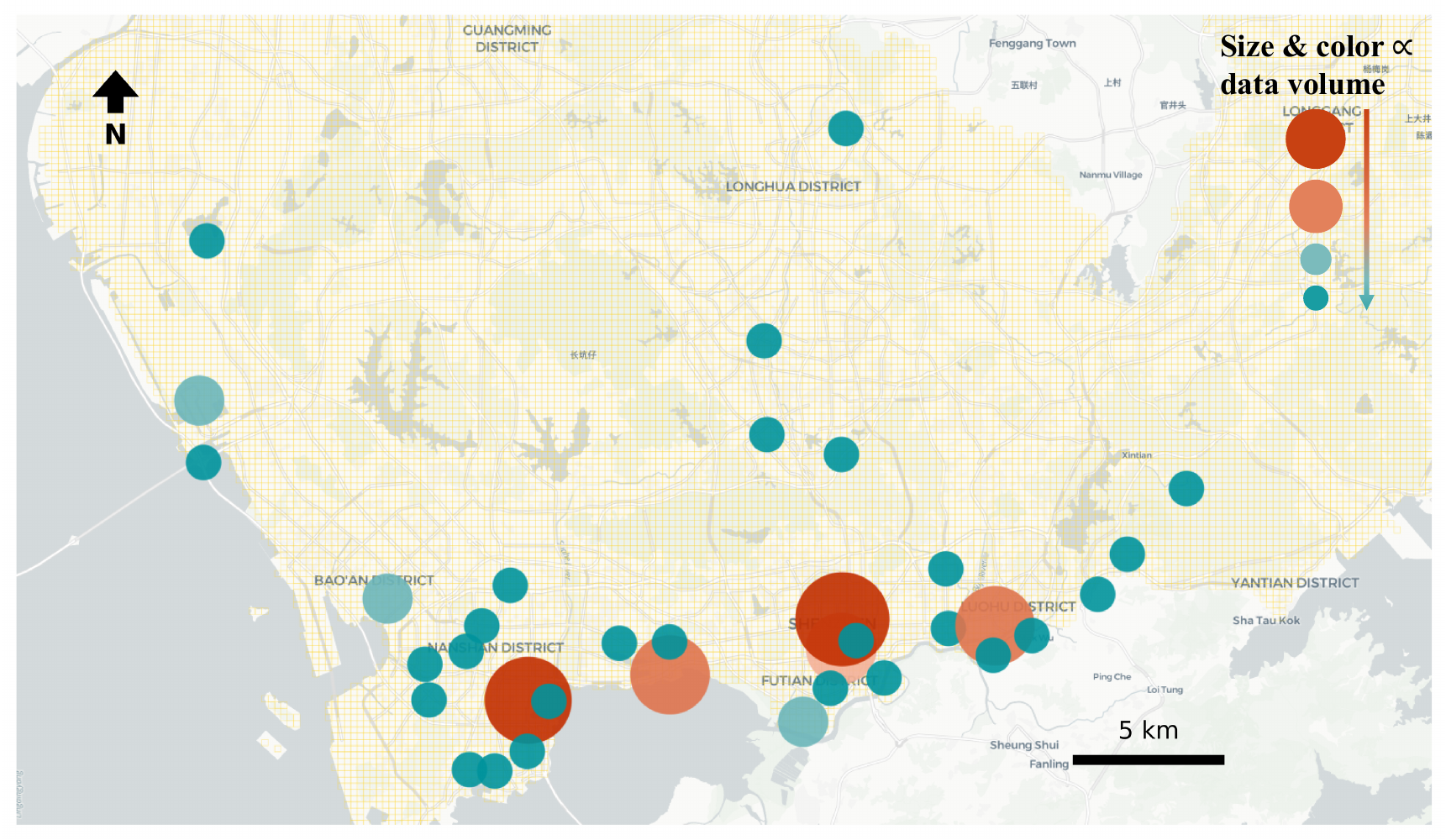}
  \caption{Spatial distribution of RedNote posts in Shenzhen.}
  \label{fig:data_dist}
\end{figure}

This dataset serves as a rich source of localized information for benchmarking how \textit{AskNearby} retrieves and recommends spatiotemporally relevant content compared to baseline LLM systems that lack grounded urban context. Furthermore, it enables fine-grained evaluation of user intent understanding, place-aware ranking, and recommendation diversity. The spatial distribution of the data is visualized in Figure~\ref{fig:data_dist}, where the heatmap highlights the concentration of activity around major commercial and cultural hubs, illustrating the high spatial resolution and urban relevance of the dataset.

\subsection{Experimental Setup}
Based on the dataset, we designed a set of local-life-oriented queries to assess the system's retrieval and generation performance.Notably, both the system and the dataset were originally in Chinese. For this paper, we provide a translated version in English to illustrate our approach. To ensure the linguistic alignment, we adopt ChatGLM-4~\cite{glm2024chatglmfamilylargelanguage}, one of the most advanced Chinese large language models available.

\subsection{Evaluation Metrics}
We use metrics covering two aspects to evaluate the retrieval and recommendation, respectively. 

For retrieval ability, we use \textbf{Precision}, \textbf{Normalized Discounted Cumulative Gain (NDCG)}, \textbf{Hallucination Rate (HR)}, and \textbf{Spatial-Temporal Relevance (STR)} to evaluate. 

\begin{itemize}
    \item \textbf{Precision}: measures the relevance of retrieved posts. High precision indicates fewer irrelevant results. Since users usually only focus on the first few results, we adopt Precision@4 in this study. 
    \item \textbf{Normalized Discounted Cumulative Gain (NDCG)}: evaluates ranking quality by incorporating graded relevance scores for each post (2 = highly relevant, 1 = moderately relevant, 0 = irrelevant). The metric computes a weighted cumulative gain for the top K results (here we use k=4) and normalizes it against the ideal ranking. 
    
    Specifically,
    \begin{equation}
        \text{NDCG}@4 = \frac{\text{DCG}@4}{\text{IDCG}@4}
    \end{equation}
    \begin{equation}
        \text{DCG}@4 = \sum_{i=1}^{4} \frac{2^{\text{rel}_i}-1}{\log_{2}(i+1)}
    \end{equation}
    \begin{equation}
        \text{IDCG}@4 = \sum_{i=1}^{4} \frac{2^{\text{rel}_i^{\star}}-1}{\log_{2}(i+1)}
    \end{equation}
    Here, $\text{rel}_i^{\star}$ represents the relevance scores in the ideal descending order. 
    \item \textbf{Hallucination Rate (HR)}: proportion of responses generated by the model with non-existent or incorrectly attributed local information (e.g., fictitious attractions or incorrect prices). 
    \item \textbf{Spatial-Temporal Relevance (STR)}: quantifies whether retrieved results are geographically constrained within the queried "proximity range" (e.g., a 1km living radius) and satisfy implicit temporal constraints (e.g., operational hours). Each result is scored 1 if both conditions are met; otherwise, it is 0.
    \item \textbf{Answer Quality (AQ)}: the overall integrity, coherence, and fluency of generated answers.
    \item \textbf{Match Score (MS)}: the degree of alignment between the answer and the query’s intent, reflecting how effectively the response resolves the user’s underlying need.
\end{itemize}

Furthermore, we introduce two LLM-evaluated metrics to complement several aspects that are hard to quantify: : (1) \textbf{Answer Quality (AQ)}: the overall integrity, coherence, and fluency of generated answers. (2) \textbf{Match Score (MS)}: the degree of alignment between the answer and the query’s intent, reflecting how effectively the response resolves the user’s underlying need. The prompts of LLMs for measuring \textbf{AQ} and \textbf{MS} are detailed in Appendix~\ref{apx:llm_prompts}.

For recommendation ability, we utilized \textbf{Hit Rate (Hit@K)}, \textbf{Normalized Discounted Cumulative Gain (NDCG@K)} and \textbf{Mean Reciprocal Rank (MRR)} to measure.

\begin{itemize}
    \item \textbf{Hit Rate (Hit@K)}: measures whether any ground-truth item appears within the top $K$ recommended items. It reflects the coverage of relevant results, regardless of position. We report Hit@5 and Hit@10 in the experiment. 
    \item \textbf{Normalized Discounted Cumulative Gain (NDCG@K)}: evaluates ranking quality by rewarding relevant items appearing earlier in the list. It is computed as described in the retrieval metrics, and we report NDCG@5 and NDCG@10. 
    \item \textbf{Mean Reciprocal Rank (MRR)}: the reciprocal of the rank at which the first relevant item appears in the recommendation list. A higher MRR indicates that relevant content is shown earlier, enhancing user satisfaction. 
\end{itemize}

These metrics effectively reflect user satisfaction in life-circle scenarios by measuring whether relevant posts appear early within a limited number of local recommendations.

Considering the nuanced judgment required by the evaluation, we employed a human evaluation methodology for the rigorous and objective comparison of our model against the baselines. The process involved three trained evaluators who assessed the outputs from all models, presented in a randomized and anonymized order to prevent bias. For relevance-based metrics like Precision@4 and NDCG@4, evaluators assigned relevance scores to each retrieved item, which were then used to compute the final values. For qualitative metrics such as Hallucination Rate (HR) and Answer Quality (AQ), they assigned scores directly based on predefined criteria including factual accuracy, spatiotemporal adherence, and contextual alignment. To ensure reliability, the final score for each metric was the average from the three evaluators, with inter-annotator agreement calculated and any significant discrepancies resolved through discussion.

\subsection{Ablation Study}
We conducted an ablation study to systematically evaluate the retrieval and recommendation capabilities of our system, respectively. This approach was chosen because our system is specifically engineered to address the novel LLIA problem by integrating dynamic user context, cognitive maps, and spatial-semantic features. Standard recommenders intrinsically lack these  capabilities, and adapting them for this task would require non-trivial modifications, creating a significant risk of comparison bias. Therefore, an ablation study provides a more direct and insightful evaluation of each component in both retrieval and recommendation modules.

Results are summarized in Table~\ref{tab:ablation_rag} and Table~\ref{tab:ablation-recommend}. For retrieval, the full three-layer RAG (GraphRAG + VectorRAG + GeoRAG) achieves the best performance. Removing GraphRAG moderately reduces Precision@4 and NDCG@4, while excluding GeoRAG leads to the largest performance drop, with STR falling from 83.8\% to 58.7\% and hallucination rate rising, underscoring its role in grounding responses. VectorRAG alone performs worst across all metrics, showing that semantic similarity retrieval is insufficient without geographic and relational constraints. Through the experiments, we have validated the necessity of combining all three retrieval strategies to ensure both contextual relevance and geographic correctness in real-world use. 

\begin{table*}[htbp]
\centering
\caption{Ablation study on different RAG-module combinations in \textit{AskNearby}.  
The first row is the full model.  Subsequent rows show the effect of removing one or more modules.}
\label{tab:ablation_rag}
\begin{tabular}{lcccccc}
\toprule
\textbf{Method} & \textbf{Precision@4 (\%)} & \textbf{NDCG@4} & \textbf{HR (\%)} & \textbf{STR (\%)} & \textbf{AQ} & \textbf{MS} \\
\midrule
\textbf{AskNearby (All RAGs)}                                      & \textbf{75.6} & \textbf{0.96} & \textbf{2.5} & \textbf{83.8} & \textbf{3.9} & \textbf{4.2} \\ \midrule
\textit{Geo + Vector  (---\,GraphRAG)}                               & 70.8 & 0.91 & 3.2 & 82.9 & 3.6 & 3.8 \\
\textit{Graph + Vector  (---\,GeoRAG)}                               & 66.1 & 0.83 & 4.5 & 58.7 & 3.4 & 3.2 \\
\textit{VectorRAG only  (---\,GeoRAG,\;---\,GraphRAG)}                 & 59.7 & 0.78 & 5.1 & 53.6 & 3.1 & 2.9 \\
\bottomrule
\end{tabular}
\end{table*}

For recommendation, combining spatial proximity, public familiarity, and semantic relevance yields the best results. Dropping semantic or public cues degrades ranking quality and early-hit accuracy, while relying on spatial proximity alone performs worst, confirming that spatial context must be complemented with social and semantic signals.

\begin{table*}[htbp]
  \small
  \centering
  \caption{Ablation results of the \textit{AskNearby} recommendation module
           (S = spatial proximity,\;P = public familiarity,\;Sem = semantic relevance).}
  \label{tab:ablation-recommend}
  \begin{tabular}{@{}lccccc@{}}
    \toprule
    \textbf{Method} & \textbf{Hit@5} & \textbf{Hit@10} & \textbf{NDCG@5} & \textbf{NDCG@10} & \textbf{MRR} \\
    \midrule
    S + P + Sem & 0.612 & 0.708 & 0.482 & 0.534 & 0.421 \\
    S + Sem       & 0.587 & 0.682 & 0.455 & 0.506 & 0.386 \\
    S + P     & 0.573 & 0.669 & 0.443 & 0.493 & 0.397 \\
    S only      & 0.542 & 0.638 & 0.412 & 0.461 & 0.357 \\
    \bottomrule
  \end{tabular}

\end{table*}

\subsection{Comparison Experiment}
We then focus on evaluating retrieval performance. Our system is compared against several representative baseline models, including both LLM-based architectures. In addition to state-of-the-art general-purpose LLMs such as GPT-4o\footnote{\url{https://chatgpt.com/}}, we also include Chinese-developed models like DeepSeek\footnote{\url{https://www.deepseek.com/}} and Qwen\footnote{\url{https://chat.qwen.ai/}}. Since these models do not inherently possess localized knowledge, we construct an external knowledge base for each and enable retrieval to ensure a fair comparison. 

To reflect real-world usability, we further compare \textit{AskNearby} with domain-specific or online map platforms, including RedNote\footnote{\url{https://www.xiaohongshu.com/}}, as well as the most widely adopted local map applications in China, such as Gaode Maps\footnote{\url{https://gaode.com/}} and Baidu Maps\footnote{\url{https://map.baidu.com/}}. These platforms have recently integrated advanced LLMs (e.g., DeepSeek-R1) to support natural language-based local information search and summarization. We conduct manual searches through the platform interface and collect the corresponding returned results. 

Here are the introductions of our baseline models:
\begin{itemize}
    \item \textbf{GPT-4o}: State-of-the-art multilingual LLMs with strong general reasoning and dialogue capabilities.
    \item \textbf{DeepSeek R1} and \textbf{Qwen3 Turbo}: High-performance Chinese LLMs optimized for local language understanding. These models demonstrate strong reasoning capabilities and complex Chinese instruction-following behavior, with significantly improved human preference alignment compared to previous generations.
    \item \textbf{RedNote}: A local lifestyle and information-sharing platform with user-generated content and keyword-based search.
    \item \textbf{Gaode Maps, Baidu Maps}: Leading map applications in China that offer location-based search with LLM-enhanced natural language interfaces.
\end{itemize}

\begin{table}[!htbp]
    \small
    \centering
    \caption{Evaluation results of \textit{AskNearby} vs. baseline models (Bold: \textbf{best}, underline: \underline{second-best})}
    \label{tab:comparision}
    \begin{tabular}{lcccccc}
        \toprule
        Method& Precision@4 $\uparrow$ (\%)& NDCG@4 $\uparrow$& HR $\downarrow$ (\%)& STR $\uparrow$ \\
        \midrule
        GPT-4o& 58.3 & 0.86 & 15.8 & 41.7 \\
        DeepSeek-R1& 72.8 & \underline{0.92} & 15.8 & \underline{64.2} \\
        Qwen-3-turbo& 70.8 & \underline{0.92} & 15.0 & 62.5 \\
        RedNote& \textbf{78.3}& 0.88& 13.1& 56.9\\
        Gaode maps& 50.0& 0.65& 9.3& 41.0\\
        Baidu maps& 65.0& 0.71& \underline{3.0}& 50.0\\
        \textbf{AskNearby (Ours)}& \underline{75.6}& \textbf{0.96}& \textbf{2.5}& \textbf{83.8}\\
        \bottomrule
    \end{tabular}
\end{table}

Table~\ref{tab:comparision} shows the evaluation results of our system. \textit{AskNearby} outperforms most baselines across retrieval metrics. However, in precision, it ranks second (75.6\%) slightly below RedNote (78.3\%). This is likely attributed to the dominance of simple keyword matching tasks in the experimental queries (e.g., "Starbucks"), where RedNote's keyword-based engine excels. In contrast, RedNote struggles with queries requiring complex intent or spatiotemporal constraints (e.g., "a quiet coffee shop for afternoon work") while \textit{AskNearby} excels with its superior semantic understanding capabilities, evident from its much higher NDCG@4 score (0.96 vs. 0.88). Notably, \textit{AskNearby} far surpasses all baselines in STR, underscoring its ability to semantically and spatiotemporally retrieve relevant results—critical in real-world local information access. In addition, \textit{AskNearby} excels in generating high-quality responses with minimal hallucinations. It effectively retrieves and summarizes relevant knowledge grounded in local context, leading to informative and trustworthy answers. 

\subsection{Qualitative Analysis}
\begin{figure}[!htbp]
  \centering
  \includegraphics[width=0.5\textwidth]{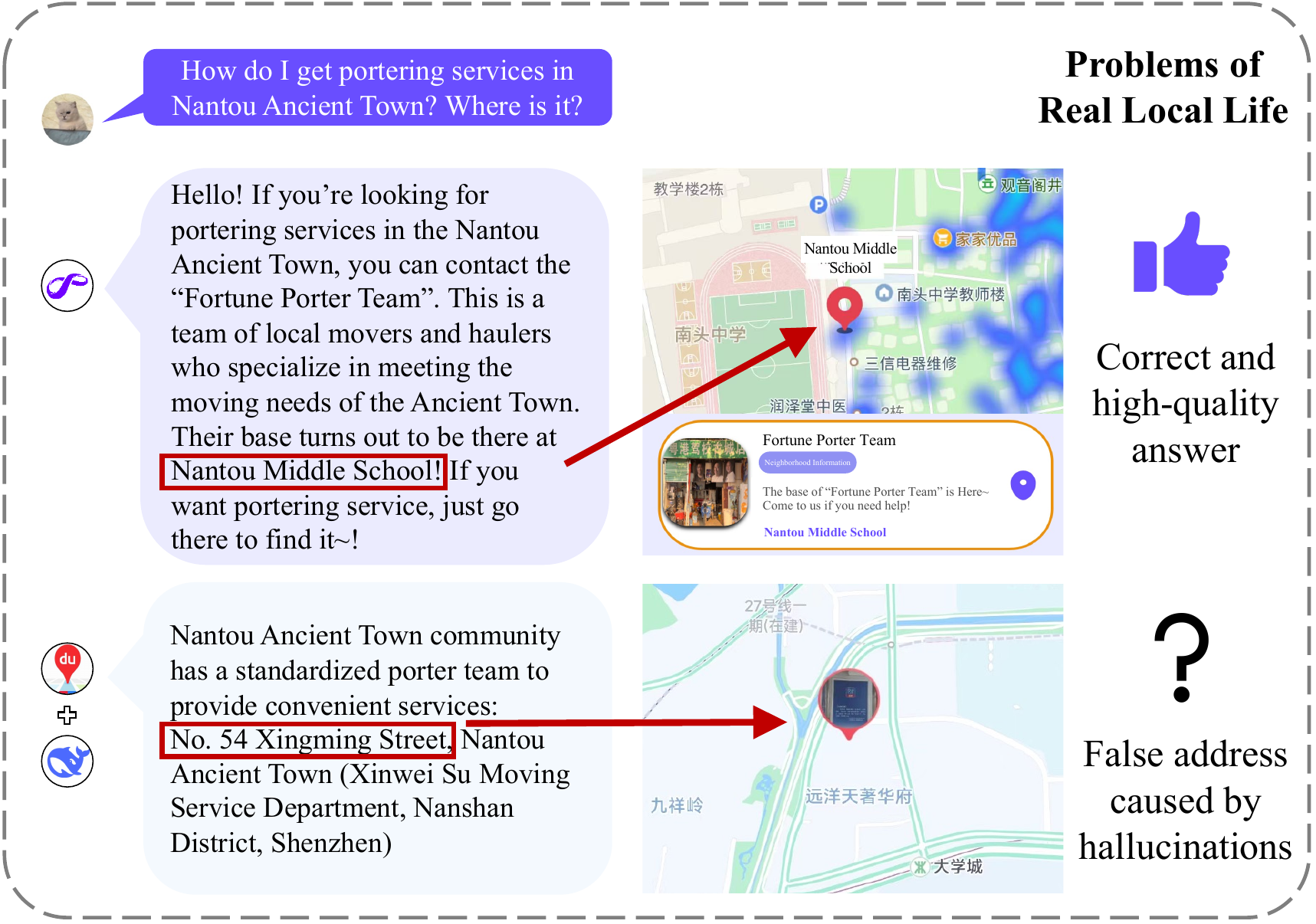}
  \caption{Answers to a local life problem from AskNearby and Baidu maps (powered by DeepSeek-R1).}
  \label{fig:quali_compare}
\end{figure}

To demonstrate the practical effectiveness of our system, we further conduct a qualitative analysis in real-world local information scenarios. Figure~\ref{fig:quali_compare} presents a representative case in which a user poses the context-specific query: "Where can I find healthcare services inside the neighborhood?". We compare the response generated by \textit{AskNearby} against a baseline LLM-based application.

\label{quali_example}

As the figure illustrates, the baseline application, while identifying semantically related content, ultimately hallucinates a non-existent address. This failure stems from a lack of accurate, localized knowledge grounding, which can mislead the user. In stark contrast, \textit{AskNearby} effectively synthesizes information from relevant posts to extract and present the correct service location, substantiated with supporting details such as contact information.

In addition to this example, generic LLMs, even when enhanced with retrieval capabilities, often suffer from hallucination when dealing with fine-grained, up-to-date local information. This further indicates the superiority of our AskNearby in satisfying the needs for the retrieval and recommendation of local information within the life circle.

\subsection{Deployed System Performance}
To validate the effectiveness and scalability of \textit{AskNearby}, we conducted extensive field deployments in two diverse real-world environments: a university campus (PKU-SZ) and a high-density, mixed-use urban community (Nantou Ancient City). Our field deployments yielded significant positive outcomes in two diverse environments. On a university campus, \textit{AskNearby} saw rapid adoption, achieving a 46\% uptake among new students within 30 days. In a dense urban community (Nantou, 38,000 residents/km\textsuperscript{2}), the system fostered significant hyperlocal engagement, as evidenced by residents initiated 4.1 queries per month on average, enabling 1,472 offline interactions.

\begin{table}[!h]
    \small
    \centering
    \caption{Deployed system performance of \textit{AskNearby} vs. baseline models (Bold: \textbf{best}, underline: \underline{second-best})}
    \begin{tabular}{lcccc}
        \toprule
        Method& AQ $\uparrow$& MS $\uparrow$ & User $\uparrow$ & Expert $\uparrow$\\
        \midrule
        GPT-4o&  3.5 & 3.8 & 3.9 & 3.1 \\
        DeepSeek-R1&  3.7 & \underline{4.0} & \underline{4.0} & \underline{4.3} \\
        Qwen-3-turbo&  3.6 & 3.7 & 3.2 & 3.3\\
        RedNote&  2.2& 3.1 & 2.7 & 2.5\\
        Gaode maps& 3.1& 3.2 & 2.8 & 3.2\\
        Baidu maps& \underline{3.8}& 3.9 & 3.7 & 3.8\\
        \textbf{AskNearby (Ours)}& \textbf{3.9}& \textbf{4.2} & \textbf{4.6} & \textbf{4.5}\\
        \bottomrule
    \end{tabular}
    \label{tab:depolyed_study}
\end{table}

Also, we conduct manual evaluation by inviting 40 regular users (\textbf{User}) and 10 domain experts (\textbf{Expert}) in urban planning to assess \textit{AskNearby} against baselines based on their individual queries. Performance was measured across four key dimensions: AQ and MS, alongside manual ratings from Users and Experts. The results, summarized in Table~\ref{tab:depolyed_study}, demonstrate the superiority of our system, confirming its effectiveness in real-world scenarios. 

This robust performance, validated in both large-scale deployments and controlled evaluations, underscores the real-world value of our approach, enhancing information accessibility and serving as a social connector that fosters neighborhood interaction.

\section{Conclusion}
In this paper, we introduce the problem of LLIA, which focuses on enabling residents to efficiently acquire relevant, timely, and context-aware information within their neighborhood. Unlike traditional search or recommendation systems that overlook fine-grained spatiotemporal dynamics and user-specific spatial cognition, LLIA highlights the importance of integrating geographic proximity, temporal constraints, and personalized perceptions in local information access. To address this challenge, we propose \textit{AskNearby}, an AI-driven community platform integrating a three-layer retrieval-augmented generation framework with a cognitive map-based recommendation model. 

Experiments on real-world datasets demonstrate that \textit{AskNearby} significantly outperforms both LLM-based baselines and existing local service platforms in retrieval accuracy, contextual alignment, and hallucination mitigation. Qualitative studies and real-world deployments further validate its effectiveness, demonstrating both accurate localized knowledge grounding and the empowerment of residents with hyperlocal knowledge. By supporting both active search and passive recommendation at the neighborhood scale, as well as the outstanding ability of understanding natural language, \textit{AskNearby} transforms static POIs into more local scenarios, allowing residents to know not only "what is there," but more importantly, "what it is like there," thereby effectively bridges the gap between physical proximity and true psychological and lifestyle accessibility. This brings the 15-minute city vision closer to reality. Moreover, unearthing fine-grained urban information overlooked by traditional maps, our work activates rich local resources, transforming communities from static geographical entities into dynamic socio-economic networks, greatly enriching the concept of the "15-minute city" and opening new possibilities for human-centric spatial computing in urban environments.

While our \textit{AskNearby} system shows promising results, we acknowledge certain limitations and identify key directions for future work. We plan to advance our research along three main fronts. First, we aim to refine the core cognitive map model, focusing on more sophisticated, adaptive mechanisms to capture individual user nuances and their evolving preferences. Second, to enhance the scientific rigor and generalizability of our findings, we will significantly expand our data foundation and refine our evaluation methodologies. This involves diversifying data sources and geographical coverage beyond initial settings, alongside exploring more robust assessment protocols. Finally, we will focus on enhancing the practical utility and real-world applicability of \textit{AskNearby}. This includes enriching its knowledge base with a wider array of information sources to improve overall accuracy and versatility, and conducting broader deployments to validate its effectiveness across diverse urban contexts and contribute to the development of intelligent localized services.

\begin{acks}
This research was supported by Dr. Qi Shu and the \textit{Shu Qi Youth Innovation Leadership Charitable Trust}. The authors would like to express sincere gratitude for their generous support and valuable guidance.
\end{acks}
\bibliographystyle{ACM-Reference-Format}
\bibliography{AskNearby_reference}

\appendix

\section{Demonstration of Our AskNearby System}
\label{apx:demonstration}
\begin{figure*}[!h]
  \centering
  \includegraphics[width=\textwidth]{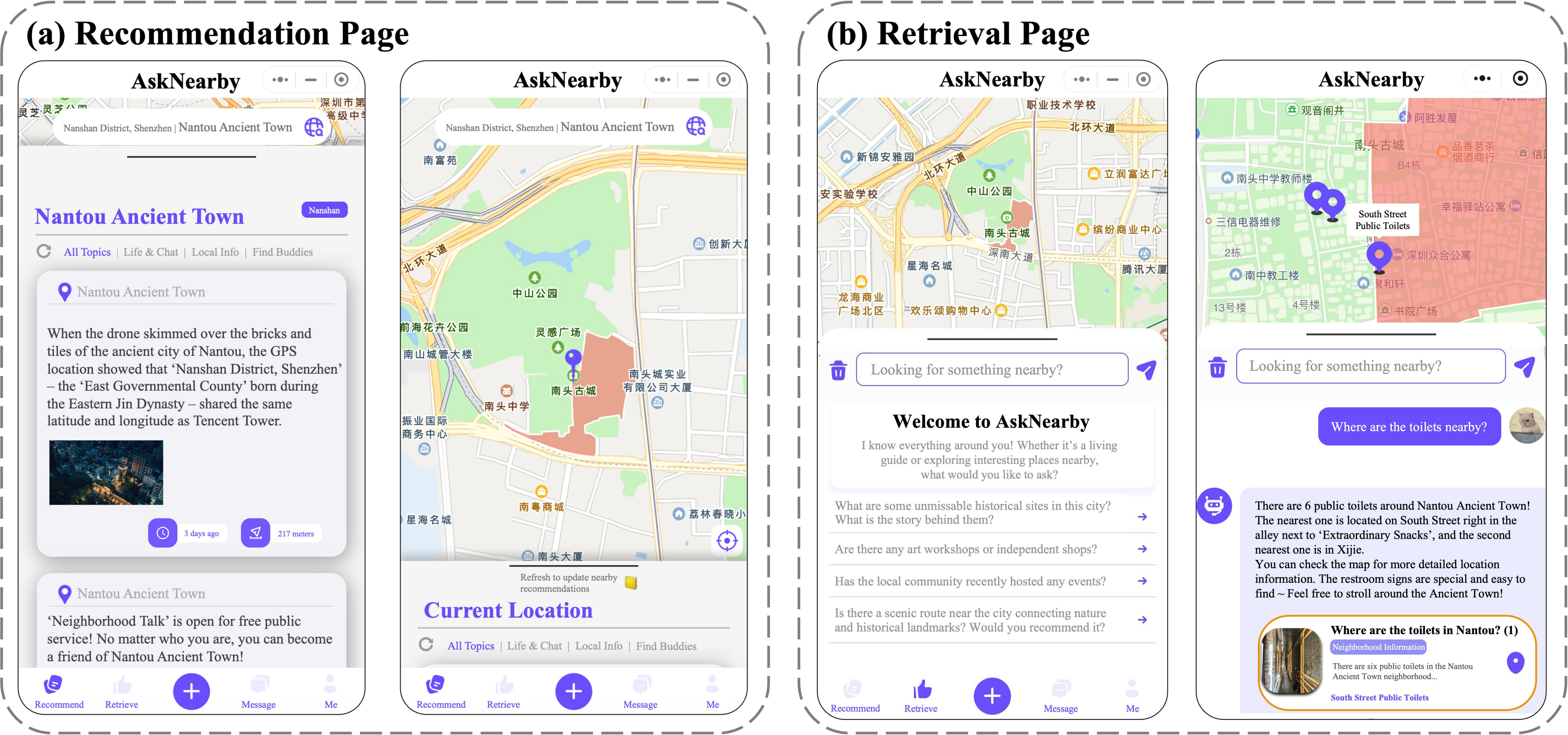}
  \caption{Screenshots of the AskNearby system: (a) Recommendation page and (b) Retrieval page.}
  \label{demonstration}
\end{figure*}

Figure~\ref{demonstration} presents the screenshots of our developed AskNearby system, which is designed to enhance local life information accessibility through both recommendation and retrieval functionalities.

Subfigure (a) illustrates the \textbf{Recommendation Page}, where users receive personalized local content suggestions based on location, time, and cognitive preferences. The left panel features a vertically scrollable list of recommended posts concerning nearby locations or events, detailed with text descriptions, timestamps, spatial distances, and multimedia content. Adjacent to this list, an interactive map interface is provided in the right panel. On this map, users can adjust the location pointer to refresh their current view and receive recommendations for different areas.

Subfigure (b) shows the \textbf{Retrieval Page}, which supports natural-language-based active querying. Users can input open-domain questions about their surroundings, such as "Where are the toilets nearby?", and the system retrieves relevant spatial entities from the spatiotemporal knowledge base. The retrieval results are displayed in textual format, providing detailed descriptions, and visually on the map, allowing users to explore spatial relations interactively. The system also suggests example queries to facilitate information discovery.

\section{LLM Evaluation Prompts}
\label{apx:llm_prompts}
\subsection{LLM Evaluation Prompt for Local Information Retrieval}
\label{apx:llm_retrieval}
To assess AQ and MS metrics, we utilized Gemini 2.5 as the evaluator, leveraging its established high correlation with human judgment. The selection of an external model, distinct from our internal ChatGLM-4 and other comparative models, was a deliberate measure to prevent self-evaluation bias. Here are the prompts.
\begin{lstlisting}
You are serving as a rigorous reviewer and need to score the following response. Please provide two integer scores on a 1-5 scale (5 being the best):

# Criteria
Answer Quality (evaluating the overall integrity, coherence, and fluency of generated answers)
Match Score (assessing the degree of alignment between the answer and the query's intent, reflecting how effectively the response addresses the user's underlying need)

# Output Format (only return JSON)
{
  "Answer Quality": <1-5>,
  "Match Score": <1-5>,
  "Comment": "<concise justification in 50 characters or less>"
}

# Input
## Question
(User's input query)
## Response
(System-generated answer)
\end{lstlisting}

\subsection{LLM Evaluation Prompt for Community Life-Circle Recommendations}
% \noindent
% The following prompt is used to evaluate life-circle recommendation results using a language model. It includes task instructions, input context, and a strict output format.
\begin{lstlisting}
You are an evaluator for a life-circle recommendation system.
Your current time is {Time}, and your location is {Location}.
Based on the recommended post list (see below), output the following evaluation as JSON:

1. How many posts are relevant to your life circle (within 5 km or a familiar neighborhood)?
2. For each post, give a match score (0.0 to 1.0) representing how useful the content is to you.
3. At which position (starting from 1) is the post that you find most interesting?

Output format (only return JSON):
{
  "relevant_post_count": <int>,
  "match_score": [<float>, <float>, ...],
  "top_interest_position": <int>
}

Post Info:
[
  {
    "sname": "Shenzhen University Town",
    "title": "Flea Market | Second-hand Bicycle for Sale",
    "post_content": "Selling a nearly new bike on campus, 80% new, sincere offer",
    "time": "2024-03-01 14:52:00",
    "latitude": 22.590,
    "longitude": 113.943,
    "tag": ["Shenzhen", "Nanshan", "Campus", "Second-hand"]
  },
  ...
]

Note: Only return valid JSON. Do not include explanations or extra text.
\end{lstlisting}
\end{document}